\documentclass[prb,twocolumn,showpacs,preprintnumbers]{revtex4}
\usepackage{amssymb}

\usepackage{graphicx}
\usepackage{dcolumn}
\usepackage{bm}

\begin{document}

\title{Mott--Hubbard metal--insulator transition at non-integer filling}
\author{Krzysztof Byczuk,$^{a,b}$ Walter Hofstetter,$^c$ and Dieter Vollhardt%
$^b$ }
\affiliation{\centerline {(a) Institute of Theoretical Physics,
Warsaw University, ul. Ho\.za 69, PL-00-681 Warszawa, Poland, }
\centerline{(b) Theoretical Physics III, Center for Electronic Correlations and Magnetism,
Institute for Physics,}\\
\centerline{University of Augsburg, D-86135 Augsburg, Germany, }\\
\centerline {(c) Lyman Physical Laboratory, Harvard University,
Cambridge, MA 02138 USA }}
\date{\today}

\begin{abstract}
Correlated electrons in a binary alloy $A_{x}B_{1-x}$ are investigated
within the Hubbard model and dynamical mean--field theory (DMFT). The random
energies $\epsilon _{i}$ have a bimodal probability distribution and an
energy separation $\Delta $.
We solve
the DMFT equations by the numerical renormalization group method at zero
temperature, and calculate the spectral density as a function of disorder
strength $\Delta $ and interaction $U$ at different fillings. For filling
factors $\nu =x$ or $1+x$ the lower or upper alloy subband is half filled
and the system becomes a Mott insulator at strong interactions, with a
correlation gap at the Fermi level. At the metal--insulator transition
hysteresis is observed. We also analyze the effective theory in the $\Delta
\rightarrow \infty $ limit and find good agreement between analytical and
numerical results for the critical interaction $U_{c}$ at which the
metal--insulator transition occurs.
\end{abstract}

\maketitle




\section{Introduction}

A Mott--Hubbard metal--insulator transition (MIT) occurs when the
local
interaction $U$ between the electrons on a lattice reaches a critical value $%
U_{c}$.\cite{Mott90,Gebhard97} Since in the Mott insulator the
particles are essentially localized on lattice sites, the
transition can only occur if the number of particles is
commensurate with the number of sites, i.e., for integer filling
factor $\nu \equiv N/N_{L}$, where $N$ and $N_{L}$ are the number
of electrons and lattice sites, respectively. If the translational
symmetry of the lattice is broken either spontaneously (as in an
antiferromagnet) or by application of some field (as in a
superlattice) the unit cell is enlarged and the Mott insulator
can, in principle, occur even at a rational filling factor $\nu
=p/q$, where $p,q\in \mathbb{N}$.\cite{Gebhard97,Miranda02,Lee03}
In fact, in our recent investigation of disordered electronic
systems \cite{byczuk03} we discovered that in binary alloy systems
$A_{x}B_{1-x}$, composed of two different atoms $A$ and $B$ with
concentrations $x$ and $1-x$, respectively, a Mott--Hubbard MIT
can even occur for arbitrary \emph{non--integer} filling factors.
This transition takes place if $%
\nu =x$ or $1+x$, where $0<x<1$. This observation extends the
traditional view of the Mott--Hubbard MIT and the notion of Mott
insulators to a wider class of systems with, basically, arbitrary
fillings. As we will argue below, in such a system magnetic
long-range order should be suppressed due to the absence of
particle--hole symmetry and, therefore, the ground state is likely
to be paramagnetic.
 Therefore, the experimental realization of such a Mott
insulator would provide an excellent playground for the study of
Mott insulators without long--range order.

In the present paper we study correlated electrons on a lattice
using the dynamical mean--field theory
(DMFT)\cite{vollhardt93,georges96} applied to the disordered
Hubbard model at zero temperature.\cite{ulmke95} The numerical
renormalization group (NRG) method is used to solve the
selfconsistent DMFT
equations.\cite{Wilson75,Costi94,bulla98,Hofstetter00} We present
results for the single--electron spectral density and the
self--energy to show how a correlated metal away from
half--filling may turn into a Mott insulator by increasing the
disorder. We find hysteresis in this transition, like the one
observed for the MIT in the pure Hubbard model at
half--filling.\cite{bulla99,rozenberg94} From the numerical data
we construct a phase diagram at $\nu =x$ and propose a new
classification scheme for correlated insulators with binary--alloy
disorder. We also develop an effective analytic theory which is
asymptotically exact in the limit of alloy band splitting, and
which shows that the Mott--Hubbard
MIT occurs at the critical interaction $U_{c}/W\approx 3\sqrt{x}/2$, where $%
W $ is the band--width of the noninteracting system.

In Ref. \onlinecite{byczuk03} we analyzed the problem of the Mott
MIT in a binary--alloy host at relatively high temperatures using
the quantum Monte--Carlo (QMC) method to solve the DMFT equations.
This only allowed us to detect a crossover from a correlated metal
to a Mott insulator. Moreover, since in Ref. \onlinecite{byczuk03}
we were primarily interested in ferromagnetism in binary--alloy
systems, we used the density of states (DOS)
corresponding to the fcc lattice in infinite dimension.\cite%
{muller-hartmann91} Such an unbounded DOS supports ferromagnetism
in a one--band Hubbard model\cite{ulmke98,vollhardt00} but does
not lead to a real gap for the Mott insulator. The NRG method
applied here is accurate enough to detect a sharp MIT in a
disordered and correlated electron system at zero temperature.
Explicit calculations were performed with a semi--elliptic Bethe
DOS, having finite support, which leads to the opening of a
genuine gap at the MIT. Additionally, the NRG allows us to
determine the self--energy on the real frequency axis and to
discuss in detail how the Mott gap opens.

The paper is organized as follows: in section II we introduce the
model and discuss the properties of interacting electrons in a
binary alloy host. In section III the DMFT method is introduced to
solve the Hubbard model with local disorder. In section IV, we
present the numerical results and provide evidence for a MIT away
from half--filling. In section V an asymptotic theory is
developed, and in section VI our conclusions are presented. In the
Appendix, we present a novel method to extract the local
self--energy within the DMFT in disordered systems.

\section{Anderson--Hubbard Hamiltonian with binary alloy disorder}

As a minimal model describing correlated lattice electrons in the
presence of disorder we consider the single--orbital
Anderson--Hubbard (AH) Hamiltonian
\begin{equation}
H_{AH}=-t\sum_{\langle ij\rangle \sigma }a_{i\sigma }^{\dagger
}a_{j\sigma }+\sum_{i\sigma }\epsilon _{i}n_{i\sigma
}+U\sum_{i}n_{i\uparrow }n_{i\downarrow },  \label{1}
\end{equation}%
where $t>0$ is the hopping integral for the electrons between
nearest neighbor sites, $U$ is the on--site interaction energy
between electrons with opposite spins $\sigma =\pm 1/2$,
$n_{i\sigma }=a_{i\sigma }^{\dagger }a_{i\sigma
}^{{\phantom{\dagger}}}$ is the local electron number operator,
and $\epsilon _{i}$ is the local ionic energy which is a random
variable. In the following we assume a bimodal probability
distribution for the random
variable $\epsilon _{i}$, i.e.,%
\begin{equation}
\mathcal{P}(\epsilon _{i})=x\delta (\epsilon _{i}+\frac{\Delta }{2}%
)+(1-x)\delta (\epsilon _{i}-\frac{\Delta }{2}),
\end{equation}%
which corresponds to a binary--alloy system composed of two
different atoms A and B. The atoms are distributed randomly on the
lattice and have ionic energies $\epsilon _{A,B}$, with $\epsilon
_{B}-\epsilon _{A}=\Delta $. The parameter $\Delta $ is a measure
of the disorder strength. The concentration of A (B) atoms is
given by $x=N_{A}/N_{L}$ ($1-x=N_{B}/N_{L}$), where $N_{A}$
($N_{B}$) is the number of the corresponding atoms.

From the localization theorem (the Hadamard--Gerschgorin theorem
in matrix algebra) it is known that if the Hamiltonian $H_{AH}$,
with a bimodal distribution for $\epsilon _{i}$, is bounded, then
there is a gap in the single--particle spectrum for sufficiently
large $\Delta \gg \max (|t|,\;U)$.\cite{Kirkpatrik70,Gonis}
 Hence at $\Delta =\Delta _{c}$ the DOS splits into two
parts corresponding to the lower and the upper alloy subbands with
centers of mass at the ionic energies $\epsilon _{A}$ and
$\epsilon _{B}$, respectively. The width of the alloy gap is of
the order of $\Delta $. The lower and upper alloy subband contains
$2xN_{L}$ and $2(1-x)N_{L}$ states, respectively. If the
Hamiltonian is \emph{not} bounded, as for example in the case of a
tight binding Hamiltonian on a hypercubic lattice in infinite
dimensions, the alloy gap is reduced to a pseudo--gap, i.e., the
spectral function vanishes only at a single point. These
statements hold for all space dimensions. However, the alloy gap
can be destroyed by clusters of one type of atoms which are
surrounded by atoms of the other type. They create an
exponentially vanishing DOS in the gap (Lifshitz tails) near the
edges of the alloy subbands.\cite{Gonis}

Binary alloy disorder in a noninteracting electron system can
create two kinds of localized states: (i) states which are
localized due to coherent backscattering processes (Anderson
localization);\cite{Anderson58,Economou70} and (ii) states in the
middle of the alloy subbands, which are localized due to a
particular superposition of the electronic wave functions caused
by particular arrangements of the alloy
atoms.\cite{Kirkpatrik70,Tong80,Naumis02} While the localized
states of type (i) are generic and gradually appear in the alloy
subbands starting from the band edges, the localized states of
type (ii) can be removed either by a small perturbation of the
ionic energies or by an interaction between the particle, and,
therefore, are beyond the scope of the present paper.

The most spectacular effect of strong correlations between the
electrons in a pure ($\Delta =0$) system is the Mott--Hubbard
metal--insulator transition. It occurs for $\nu =1,2,\dots ,2g-1$,
where $g$ is an orbital degeneracy, at an interaction $U=U_{c}$
(note that $\nu =2g$ corresponds to a band
insulator).\cite{rozenberg94,rozenberg92,zhang93,Kotliar96,kajuter97}
 For interactions $U\ll U_{c}$ the electrons gain kinetic energy due
to the delocalized nature of the wave function. When $U$
increases, the electrons keep at a distance as much as possible,
and at $U\geq U_{c}$ the many--body wave function is essentially
localized at each lattice site. The system is then a Mott
insulator. Within the DMFT for Hubbard--like models with single or
degenerate orbitals a Mott--Hubbard MIT was shown to occur at
integer filling factors $\nu $.\cite{Kotliar96,kajuter97}
 At low but finite
temperatures this MIT is discontinuous, while at zero temperature
it is continuous, i.e., when approaching the critical interaction
from the metallic side the quasiparticle peak continuously narrows
until it completely disappears at the transition
point.\cite{georges96,bulla99,rozenberg94} In the insulating phase
the DOS is zero at the Fermi level and the whole spectral weight
is shifted into the two Hubbard subbands, which are remnants of
the atomic levels with single and double (multiple) occupancy.
Away from integer filling the system is always metallic and the
spectral function has visible Hubbard subbands at any finite and
large $U$.

\begin{figure}[tbp]
\includegraphics [clip,width=7.cm,angle=-00]{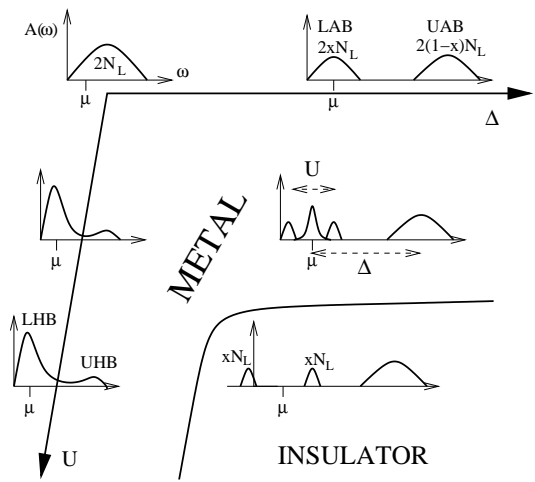}
\caption{Schematic plot representing the Mott--Hubbard
metal--insulator transition in a correlated electron system with
the binary alloy disorder. The shapes of spectral functions
$A(\protect\omega)$ are shown for different interactions $U$ and
disorder strengths $\Delta$. Increasing $\Delta$ at $U=0$ leads to
splitting of the spectral function into the lower (LAB) and the
upper (UAB) alloy subbands, which contain $2xN_L$ and $2(1-x)N_L$
states respectively. Increasing $U$ at $\Delta=0$ leads to the
occurrence of lower
(LHB) and upper (UHB) Hubbard subbands. The Fermi energy for filling $%
\protect\nu=x$ is indicated by $\protect\mu$. At $\protect\nu=x$ (or $%
\protect\nu=1+x$, not shown in the plot) the LAB (UAB) is
half--filled. In this case an increase of $U$ and $\Delta$ leads
to the opening of a correlation gap at the Fermi level and the
system becomes a Mott insulator. } \label{fig0}
\end{figure}

New possibilities appear in systems with correlated electrons and
binary alloy disorder. The Mott--Hubbard metal insulator
transition can occur at any filling $\nu =x$ or $1+x$,
corresponding to a half--filled lower or to a
half--filled upper alloy subband, respectively, as shown schematically for $%
\nu =x$ in Fig.~(\ref{fig0}). The Mott insulator can then be
approached either by increasing $U$ when $\Delta \geq \Delta _{c}$
(alloy band splitting limit), or by increasing $\Delta $ when
$U\geq U_{c}$ (Hubbard band splitting limit). The nature of the
Mott insulator in the binary alloy system can be understood
physically as follows. Due to the high energy cost of the order of
$U$ the randomly distributed ions with lower (higher) local
energies $\epsilon _{i}$ are singly occupied at $\nu =x$ ($\nu
=1+x$), i.e., the double occupancy is suppressed. In the Mott
insulator with $\nu =x$ the ions with higher local energies are
empty and do not contribute to the low--energy processes in the
system. Likewise, in the Mott insulator with $\nu =1+x$ the ions
with lower local energies are double occupied implying that the
lower alloy subband is blocked and does not play any role. We note
that for finite $U$ \emph{virtual} processes leading to double
occupation either in the lower ($\nu =x$) or in the upper alloy subband ($%
\nu =1+x$) are still possible, leading to the antiferromagnetic
superexchange interaction. However, since the positions of the
corresponding atoms are random, particle--hole symmetry is absent
such that long--range antiferromagnetic order is expected to be
suppressed. Of course, antiferromagnetism cannot be ruled out
completely on these reasoning and its appearance in the model
requires further studies.

\begin{figure}[tbp]
\includegraphics [clip,width=7.cm,angle=-00]{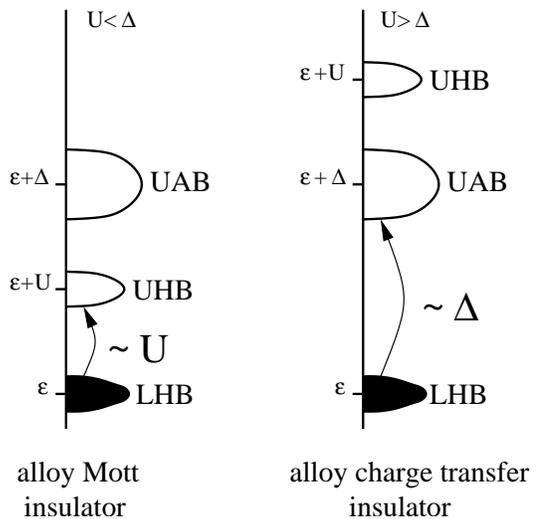}
\caption{Two possible insulating states in the correlated electron
system with binary--alloy disorder. When $U<\Delta$ the insulating
state is an alloy Mott insulator with an excitation gap in the spectrum of the order of $%
U$. When $U>\Delta$ the insulating state is an alloy charge
transfer insulator with an excitation gap of the order of
$\Delta$. } \label{fig8}
\end{figure}

For $U>U_{c}(\Delta )$ in the Mott insulating state with binary
alloy disorder one may use the lowest excitation energies to
distinguish two different types of insulators. Namely, for
$U<\Delta $ an excitation must overcome the energy gap between the
lower and the upper Hubbard subbands, as indicated in
Fig.~(\ref{fig8}). We call this insulating state an \emph{alloy
Mott insulator}. On the other hand, for $\Delta < U$ an excitation
must overcome the energy gap between the lower Hubbard subband and
the upper alloy--subband, as shown in Fig.~(\ref{fig8}). We call
this insulating state an \emph{alloy charge transfer insulator}.

\section{Dynamical mean--field theory for the disordered Hubbard model}

The Mott--Hubbard MIT is driven by the interaction between the
particles. Since this transition occurs when the interaction
energy is comparable with the single--particle energy of the
electrons, there is no natural small parameter (e.g. $t$ or $U$)
in the theory. The problem is generically nonperturbative.
Moreover, when the transition appears between a paramagnetic metal
and a paramagnetic insulator, there is no obvious order parameter
characterizing the insulating phase. In the following the
insulating state at $T=0$ is defined by the vanishing of the
one--particle spectral function at the Fermi level.

The model defined by the Hamiltonian (\ref{1}) is not exactly
solvable for any finite number of space dimensions. However, with
a proper rescaling of the hopping integral it becomes numerically
solvable in infinite dimensions, i.e., within
DMFT.\cite{metzner89}$^,$\cite{georges96,vollhardt93} For
finite--dimensional systems DMFT is a self--consistent
approximation scheme which takes into account local quantum
fluctuations but neglects spatial correlations. Since DMFT is a
nonperturbative method, it is ideally suited to study the
Mott--Hubbard MIT.

To derive the DMFT equations for the problem at hand we select a
single lattice site, $i$, and integrate out all the electronic
degrees of freedom corresponding to other sites.\cite{georges96}
 This leads to an effective single--impurity Anderson
Hamiltonian
\begin{eqnarray}
H^{\mathrm{SIAM}} &=&\sum_{\sigma }(\epsilon _{i}-\mu )a_{i\sigma
}^{\dagger
}a_{i\sigma }+Un_{i\uparrow }n_{i\downarrow } \\
&&+\sum_{\mathbf{k}\sigma }V_{\mathbf{k}}a_{i\sigma }^{\dagger }c_{\mathbf{k}%
\sigma }+V_{\mathbf{k}}^{\ast }c_{\mathbf{k}\sigma }^{\dagger
}a_{i\sigma }+\sum_{\mathbf{k}\sigma }\epsilon
_{\mathbf{k}}c_{\mathbf{k}\sigma }^{\dagger }c_{\mathbf{k}\sigma
},  \nonumber  \label{2}
\end{eqnarray}%
where $\mu $ is the chemical potential, $V_{\mathbf{k}}$ and $\epsilon _{%
\mathbf{k}}$ are the hybridization matrix element and the
dispersion relation for the auxiliary bath fermions
$c_{\mathbf{k}\sigma }$, respectively. In the present paper the
Hamiltonian (\ref{2}) is solved at
zero temperature using the numerical renormalization group method.\cite%
{Wilson75,Costi94,bulla98,Hofstetter00} For each ionic energy
$\epsilon _{i}$ we obtain the local Green function $G(\omega
,\epsilon _{i})$. The physical Green function (\ref{1}) is
obtained by algebraic averaging of $G(\omega ,\epsilon _{i})$ over
different realizations of the disorder,\cite{ulmke95} i.e.,
\begin{equation}
G(\omega )=\int d\epsilon _{i}\mathcal{P}(\epsilon _{i})G(\omega
,\epsilon _{i}).  \label{3}
\end{equation}%
From the $\mathbf{k}$-integrated Dyson equation
\begin{equation}
G^{-1}(\omega )=\omega -\eta (\omega )-\Sigma (\omega )  \label{4}
\end{equation}%
we determine the local self--energy $\Sigma (\omega )$. The
function $\eta (\omega )$, called hybridization function, is
defined as
\begin{equation}
\eta (\omega )=\sum_{\mathbf{k}}\frac{|V_{\mathbf{k}}|^{2}}{\omega
-\epsilon _{\mathbf{k}}}.  \label{5}
\end{equation}%
The DMFT equations are closed by a Hilbert transform, relating the
local Green function for a given crystallographic lattice to the
self--energy, i.e.,
\begin{equation}
G(\omega )=\int d\epsilon \frac{N_{0}(\epsilon )}{\omega -\epsilon
-\Sigma (\omega )},  \label{6}
\end{equation}%
where $N_{0}(\epsilon )$ is the non--interacting DOS. Eqs.
(\ref{2}--\ref{6}) constitute a closed set of DMFT equations for
the disordered Hubbard model.

We solve Eqs. (\ref{2}--\ref{6}) for a Bethe lattice with infinite
connectivity. In this case the DOS is given by
\begin{equation}
N_{0}(\epsilon )=\frac{2}{\pi D}\sqrt{D^{2}-\epsilon ^{2}},
\label{7}
\end{equation}%
with the bandwidth $W=2D$. In the following we set $D=1$. With the DOS (\ref%
{7}) the Hilbert transform (\ref{6}) can be calculated
analytically leading to a simple algebraic relation between the
local Green function $G(\omega )$ and the hybridization function
$\eta (\omega )$, i.e.,
\begin{equation}
\eta (\omega )=\frac{D^{2}}{4}G(\omega ).
\end{equation}

In the numerical calculations we adjust the value of the chemical potential $%
\mu $ so that the number of particles in the system is fixed.
Hence the independent variables are $\nu $, $U$, $x$, and $\Delta
$.

Since DMFT neglects short--range spatial correlations, and hence
does not include effects due to backscattering of electrons on the
randomly distributed ions, it cannot describe effects of Anderson
localization. On the other hand, binary alloy disorder is a
particularly strong type of disorder since it even leads to band
splitting -- and thereby to insulating behavior --  in any spatial
dimension. This dominant feature, and all other effects caused by
the simultaneous presence of interactions and disorder, is well
captured by DMFT. In particular, the DMFT equations reduce to the
equations of the coherent potential approximation (CPA) for
interaction $U=0$.\cite{vlaming92}
 The CPA method is known to
be very successful in explaining single-particle properties of
disordered systems, both in the case of models and realistic
systems.\cite{Gonis,Koepernik} In particular, it reproduces the
alloy band splitting in binary alloy systems. Therefore we use the
DMFT to describe the Mott--Hubbard MIT in the presence of binary
alloy disorder, and consider additional effects due to Anderson
localization as secondary.

\section{Numerical results for the disordered Hubbard model}

In the following we present our numerical results for filling
factors $\nu \neq x$ and $\nu =x$ with equal concentration of A
and B atoms, i.e., $x=1/2$. In particular, the ground-state phase
diagram at $\nu=x$ is presented and the MIT is discussed in
details.

\subsection{Interacting electrons in the alloy band splitting limit for
$\protect\nu\neq x $}

\begin{figure}[tbp]
\includegraphics
[clip,width=7.cm,angle=-00]{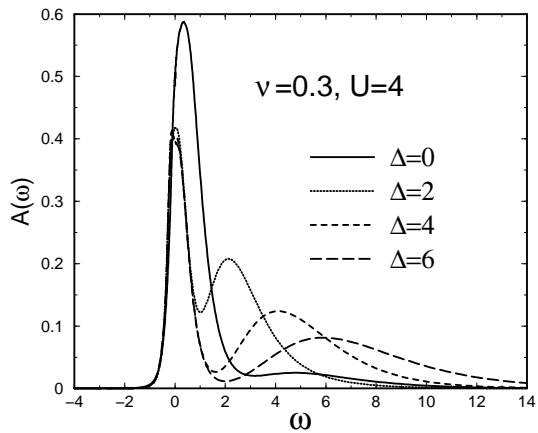}
\caption{Spectral function of the Hubbard model at $U=4$ and $\protect\nu%
=0.3 $ for different binary alloy disorder strengths $\Delta$. The
upper alloy subband splits off for $\Delta\geq 2$. The shape of
the lower alloy subband is not significantly changed for large
$\Delta$.} \label{fig1a}
\end{figure}

The influence of the disorder--induced alloy band splitting on the
spectral function $A(\omega )$ is shown in Fig.~(\ref{fig1a}) for
$U=4$ and filling factor $\nu =0.3$. For vanishing disorder
($\Delta =0)$ the spectral function is composed of the lower
Hubbard subband (which at this low density
is merged with the quasiparticle peak) and the upper Hubbard subband around $%
\omega =4$. Upon increasing $\Delta $ the upper alloy subband
splits off and moves to larger $\omega $. At the same time the
lower alloy subband appears with a smaller number of states. Its
shape and the position with respect to the Fermi energy does not
change for $\Delta >2$. These results and the shape invariance of
the lower alloy subband in the presence of interactions between
the electrons suggest that in the $\Delta \rightarrow \infty $
limit the binary alloy disordered Hubbard model can be effectively
described by the Hubbard model in a reduced Hilbert space which
contains only the states from the lower alloy subband.

\subsection{Phase diagram and Mott--Hubbard transition at filling $\protect\nu =x$}

\begin{figure}[tbp]
\includegraphics
[clip,width=7.cm,angle=-00]{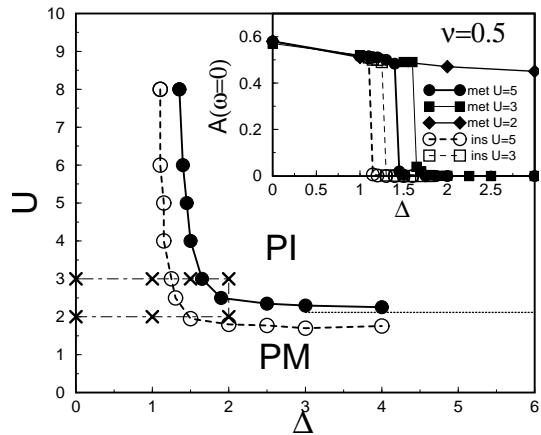}
\caption{Ground state phase diagram of the Hubbard model with
binary--alloy disorder at filling $\protect\nu=x=0.5$. The filled
(open) dots represent the boundary between paramagnetic metallic
(PM) and paramagnetic insulating (PI) phases as determined by DMFT
with the initial input given by the metallic (insulating)
hybridization function. The horizontal dotted line represents
$U_c$ obtained analytically from an asymptotic theory in the limit
$\Delta \to \infty$ (see section V). Stars show the points at
which the explicit spectral functions are presented in
Figs.~\ref{fig1bis} -- \ref{fig3bis}. Inset: hysteresis in the
spectral functions at the Fermi level obtained from DMFT with an
initial metallic (insulating) host represented by filled (open)
symbols and  solid (dashed) lines.} \label{fig7}
\end{figure}

By solving numerically the DMFT equations we extracted the
 zero temperature phase diagram
 at $\nu =x=0.5$ which is presented in Fig.~\ref{fig7}. The curve with filled
dots represents the critical interaction $U_{c2}=U_{c2}(\Delta )$
separating the paramagnetic metal (PM) and the paramagnetic
insulator (PI). This boundary line was determined by solving
iteratively the DMFT equations using a {\em metallic}
hybridization function as an initial input. It means that in
solving the system of Eqs.  (\ref{2}--\ref{6}) iteratively we
began with the hybridization function $\eta^{(0)}(\omega)\approx
G^{(0)}(\omega)$ that has a nonvanishing imaginary part at
$\omega=0$. The other curve (open dots) represents the boundary
$U_{c1}=U_{c1}(\Delta )$ between the metallic and insulating
phases, as determined by solving the DMFT equations with an {\em
insulating} hybridization function as an initial input. In this
case the initial hybridization function had vanishing imaginary
part at $\omega=0$. The boundary points (lines) correspond to the
values of the $(U,\Delta)$ parameters where the converged spectral
functions obtained from Eqs. (\ref{2}--\ref{6}) starts to have
zero weight at $\omega=0$. In the inset to Fig.~\ref{fig7}, the
behavior of the spectral function at the Fermi level is shown when
the metal--insulator boundary is approached from the metallic
(solid lines) and from the insulating side (dashed lines),
respectively. We observe hysteresis, which indicates that at low
but finite temperatures the MIT transition in the Hubbard model
with binary alloy disorder is discontinuous. By contrast, at zero
temperature the transition should be continuous and occurring at
the $U_{c}=U_{c2}(\Delta )$ line.\cite{georges96,bulla99} We
calculated numerically the average density of double occupied
sites $d=\langle n_{\uparrow}n_{\downarrow} \rangle$ in the
coexistence regime and found that $d$ is larger for a metallic
solution. It implies that the metallic ground state is
energetically more stable.\cite{georges96} Of course, very close
to $U_{c2}(\Delta)$ we cannot make an absolutely precise statement
because of the finite numerical accuracy.

From the inset to Fig.~\ref{fig7} we also conclude that in the metallic phase $%
A(0)$, the spectral function at the Fermi level, decreases with
disorder but remains independent of $U$. This behavior corresponds
to the ``pinning'' of the spectral function (Friedel sum rule) in
the pure case, where $A(0)$ does not depend on the interaction $U$
between the electrons.\cite{muller89}
 Similar behavior is encountered in
the disordered Hubbard model studied here. However, now $A(0)$ is
reduced by the $\Delta $ dependent imaginary part of the
self--energy.

\begin{figure}[tbp]
\includegraphics
[clip,width=8.cm,angle=180]{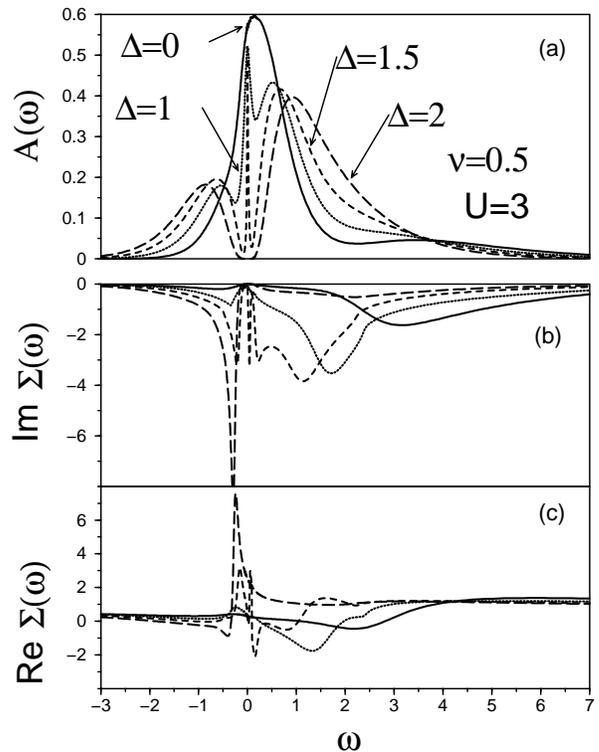}
\caption{(a) Spectral function, (b) imaginary part of the
self--energy, and (c) real part of the self--energy for the
Hubbard model at $\nu=0.5$, $U=3$ and different disorder $\Delta$.
As $\Delta$ increases, the quasiparticle peak appears and then
vanishes, signaling a transition from a metallic to an insulating
phase. In the insulating phase ($\Delta=2$) a Mott gap is opened
at $\omega=0$ with $\mathrm{Im}\Sigma(0)=0$ whereas
$\mathrm{Re}\Sigma(0)$ remains finite. } \label{fig1bis}
\end{figure}

\begin{figure}[tbp]
\includegraphics
[clip,width=8.cm,angle=180]{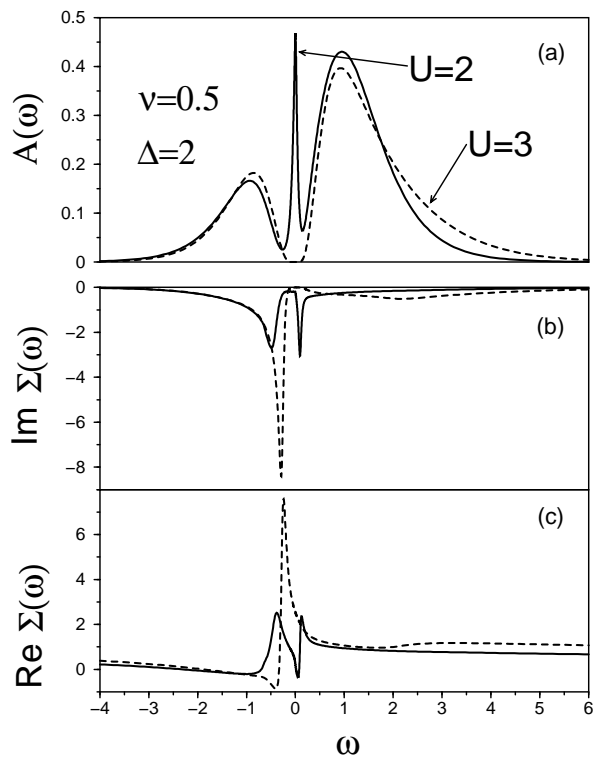}
\caption{(a) Spectral function, (b) imaginary part of the
self--energy, and (c) real part of the self--energy for the
Hubbard model at $\Delta=2$ and
 $U=3$ and $2$.
} \label{fig2bis}
\end{figure}

\begin{figure}[tbp]
\includegraphics
[clip,width=8.cm,angle=180]{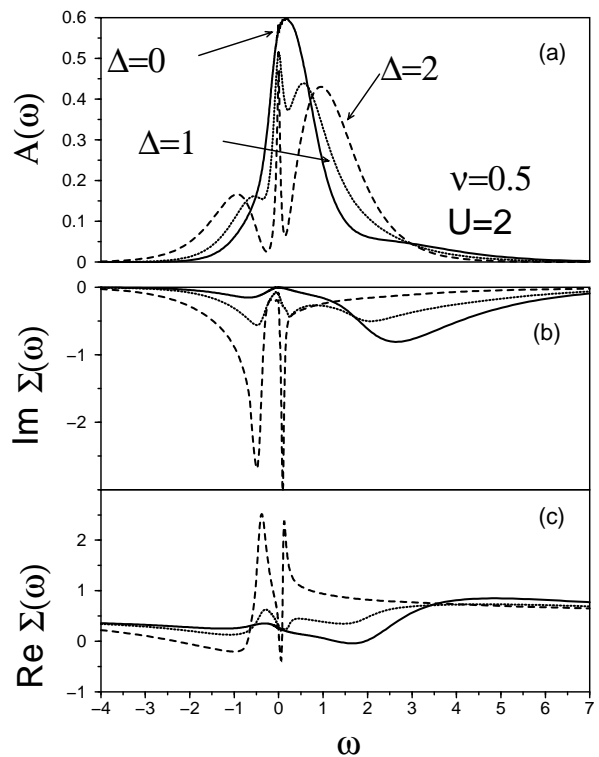}
\caption{(a) Spectral function, (b) imaginary part of the
self--energy, and (c) real part of the self--energy for the
Hubbard model at $U=2$ and different disorder $\Delta$. Upon
increasing $\Delta $ the alloy subbands are split and a
quasiparticle resonance emerges between the Hubbard subbands. }
\label{fig3bis}
\end{figure}

In the upper panels of Figs.~\ref{fig1bis}-\ref{fig3bis} we
present the spectral functions for selected parameters $U$ and
$\Delta$ along a path in the $(U,\Delta)$ phase diagram
(Fig.~\ref{fig7}) indicated by crosses: $U=3$ and
$\Delta=0\rightarrow 1 \rightarrow 1.5 \rightarrow 2$
(Fig.~\ref{fig1bis}), $\Delta=2$ and $U=3\rightarrow 2$
(Fig.~\ref{fig2bis}), and finally, $U=2$ and $\Delta=2\rightarrow
1 \rightarrow 0$ (Fig.~\ref{fig3bis}). These spectral functions
illustrate the evolution of the system within, or between, a
metallic and an insulating phase when disorder and interaction are
changed. In particular, in Fig.~\ref{fig1bis} we see for $U=3$ how
increasing $\Delta$ from the value $0$ (where the spectral
function is qualitatively similar to that of the $\nu=0.3$ case,
c.f. Fig.~\ref{fig1a}) to $1$ and $1.5$ leads to the splitting of
the alloy subbands and the emergence of the quasiparticle peak at
$\omega=0$, a feature of strong correlations between the
electrons. At $\Delta=2$ the quasiparticle peak is absent and the
spectral function possesses a Mott gap at the Fermi level, a
feature of an insulator. Keeping $\Delta=2$ in Fig.~\ref{fig2bis}
and lowering $U$ from $3$ to $2$ leads to a shrinking of the Mott
gap and reappearance of a quasiparticle peak, characterizing a
correlated metallic phase. Finally, upon lowering $\Delta$ from
$2$ to $0$ at constant $U=2$ the alloy subbands approach each
other and the quasiparticle peak merges with the lower Hubbard
subband as presented in Fig.~\ref{fig3bis}.

In addition to the spectral functions, the imaginary and real
parts of the self-energy, calculated by the method presented in
the Appendix, are shown in panels (b) and (c) in
Figs.~\ref{fig1bis}--\ref{fig3bis}. In the metallic phase with
$\Delta>0$, two important features of the self-energy should be
noted: (i) the imaginary part of the self-energy at the Fermi
level is finite, i.e., $\mathrm{Im} \Sigma(\omega=0)<0$, and (ii)
the real part of the self-energy at the Fermi level has a negative
slope, i.e., $\partial \mathrm{Re}\Sigma (\omega =0)/\partial
\omega <0$. While the former feature (caused by disorder even in
the presence of the local interaction) can be observed within the
perturbation expansion with respect to small $\Delta$, the latter
is surprising since at $U=0$ the slope is always positive for any
$\Delta>0$.\cite{laad01}

At the MIT the behavior of the self-energy changes. In particular,
the imaginary part becomes vanishingly small at $\omega=0$ whereas
the real part is finite. These results imply that the mechanism
for opening a correlation (Mott) gap at this MIT transition is
different from that in the pure Hubbard model with particle-hole
symmetry.\cite{georges96,bulla99} Namely, consider the spectral
function at the Fermi level, which
 is expressed in terms of the self--energy:
\begin{equation}
A(0)=-\frac{1}{\pi}\int d\epsilon N_0(\epsilon)
\frac{\mathrm{Im}\Sigma (0)}{ [\epsilon - \mathrm{Re}\Sigma (0)]^2
+ [\mathrm{Im}\Sigma (0)]^2}.
\end{equation}
Since we obtained numerically that at the Fermi level the
imaginary part of the self--energy vanishes and the real part is
larger then the band--width $W$, we find
$A(0)=N_0[\mathrm{Re}\Sigma (0)]=0$. This result is in contrast to
the Mott--Hubbard MIT in the pure Hubbard model at half--filling
with particle--hole symmetry. In this last case the opening of a
correlation gap at $T=0$ is due to the formation of a delta--like
singularity in the imaginary part of the self--energy at the Fermi
level. The real part of $\Sigma(\omega)$ has a $1/\omega$
divergence at this point, consistent with the Kramers--Kronig
relations. In the disordered case we do not see such a behavior of
$\mathrm{Re}\Sigma (\omega)$, which implies that lifting of the
particle--hole symmetry due to finite disorder and noninteger
filling affects the way how the gap is opened in the Mott
insulator.

\section{Asymptotic limit $\Delta \to \infty$}

Our understanding of the Hubbard model with binary alloy disorder
is based
on the fact that the lower and upper alloy subbands are split at large $%
\Delta$. We now show that for $\Delta\rightarrow \infty$ the upper
(or lower) alloy subband can be neglected and the problem can be
mapped onto a low--energy subspace of the full Hilbert space.
Effectively, the correlated and binary--alloy disordered
electronic system is represented by a correlated {\em pure}
system with renormalized parameters.

\subsection{Mapping of the Hilbert space}

We consider the case $\Delta \gg \max (U,W)$. Then the Hilbert
space can be
divided into two subspaces A and B consisting of ions with energies $%
\epsilon_A$ and $\epsilon_B$, respectively. We denote the
projection operator onto the A--subspace by $\hat{P}$ and the
projection operator onto the B--subspace by
$\hat{Q}=1-\hat{P}$.\cite{spalek,projection} The Schr\"odinger equation
can be decomposed as
\begin{eqnarray}
(E-\hat{P}H\hat{P})\hat{P}|\Psi\rangle = \hat{P}H\hat{Q} \hat{Q}
|\Psi
\rangle  \nonumber \\
(E-\hat{Q}H\hat{Q})\hat{Q}|\Psi\rangle = \hat{Q}H\hat{P} \hat{P}
|\Psi \rangle,
\end{eqnarray}
where $|\Psi\rangle$ is a many--body eigenstate of the Hamiltonian
(\ref{1}) with eigenvalue $E$. Solving this set of equations, the
effective Hamiltonian of the lower alloy subband is formally given
by
\begin{equation}
H_{\mathrm{eff}}=\hat{P}H\hat{P} - \hat{P}H\hat{Q}\frac{1}{\hat{Q}H\hat{Q}}%
\hat{Q}H\hat{P}.
\end{equation}
The mapping implies that the number of lattice sites corresponding
to the A--subspace is equal to the number of $\epsilon_A$ ions,
i.e., $N_A=xN_L$.
Therefore, if we restrict ourselves to this low--energy many--body subspace $%
\hat{P} |\Psi \rangle$, and consider only the effective Hamiltonian $H_{%
\mathrm{eff}}$, the filling factor can be replaced by
$\nu^*=\nu/x$.

The interaction $U$ is not changed by the projection of the
Hilbert space because it is a local quantity. However, the
band--width of the lower alloy subband is renormalized because the
number of nearest neighbors with onsite energy $\epsilon_A$ is
reduced. In order to estimate how the band--width renormalizes we
consider the Bethe lattice with a finite coordination number $z$
and calculate the second moment $\mu^{(2)}_i=\langle
A|H_0^2|A\rangle$ for the non--interacting Hamiltonian
$H_0=t\sum_{i,\delta} a^{\dagger}_{i} a_{i+\delta}$ with nearest
neighbor hopping. For a given lattice site $A$ we
can have $z_A$ neighbors with energies $\epsilon_A$, where $0\leq z_A \leq z$%
. It is easy to show that the second moment at site $A$ is
$\mu^{(2)}_i(z_A)=z_At^2$. The probability distribution of $z_A$
is given by
\begin{equation}
\mathcal{P}(z_A)=\left(
\begin{array}{c}
z \\
z_A%
\end{array}
\right) x^{z_A}(1-x)^{z-z_A}.
\end{equation}
Therefore we find the average moment $\langle
\mu^{(2)}\rangle$:\cite{Naumis02}
\begin{equation}
\langle \mu^{(2)} \rangle = \frac{1}{N_A}\sum_{i=1}^{N_A}
\mu_{i_A}^2(z_A) \mathcal{P}(z_A) = zt^2 x.
\end{equation}
For a Bethe lattice with coordination number $z$ tending to
infinity we rescale $t\rightarrow t^*=t/\sqrt{z}$. We then find
that the second moment is $\langle \mu^{(2)} \rangle= t^{*2} x$.
It means that the band--width, as measured by the second moment,
is reduced in the effective Hamiltonian $H_{\mathrm{eff}}$ by the
factor $\sqrt{x}$.

\subsection{Critical interaction $U_c$}

The approximate value of the critical interaction for the
occurrence of the Mott transition can be found analytically within
the linearized DMFT, where the full DMFT problem is mapped onto
the two--site SIAM with self--consistency
conditions.\cite{Bulla00} For the pure Hubbard model it was shown
that the critical interaction has the value
$U_c=6\sqrt{\mu^{(2)}}$. From the results in the last subsection
we find that the critical interaction for the MIT in the
strong disorder limit $\Delta\rightarrow \infty$ is given by $U_c=6t^*\sqrt{x%
}$. For parameters used in our calculations $t^*=0.5$ and $x=0.5$ we obtain $%
U_c=3/\sqrt{2}$. This value is shown by the dotted line in Fig.~(\ref{fig7}%
). The agreement between our numerical calculation and this
estimate of $U_c$ is surprisingly good.

\section{Conclusions}

In this paper we studied the Mott--Hubbard transition in a
correlated electronic system with binary alloy disorder. By
numerically solving the DMFT equations at $T=0$ we showed that for
filling factor $\nu=x$ a MIT occurs when the disorder $\Delta$ or
the interaction $U$ are increased. Regarding the excitation
spectrum for the electrons we introduced the notion of the "alloy
Mott insulator" for $U<\Delta$, and of the "alloy charge transfer
insulator" for $U>\Delta$. This classification is analogous to the
Zaanen--Sawatzky--Allen scheme for two band--Hubbard
systems.\cite{ZSA} In our case however, the role of the oxygen
band is played by the upper alloy subband. We also found
hysteresis upon approaching the metal--insulator boundary,
depending on the initial conditions imposed in the iterative
solution of the DMFT equations. It shows that hysteresis is a
generic feature of the MIT in pure and in binary alloy disordered
systems within the
 DMFT scenario.
We also found that the opening of a Mott gap is associated with
the disappearance of the imaginary part of the self-energy at the
Fermi level.
 Finally we discussed the analytical theory, valid in the
alloy band--splitting limit, and showed that the Hubbard model
with binary alloy disorder can be mapped onto an effective Hubbard
Hamiltonian with renormalized bandwidth and filling factor. The
estimated critical interaction $U_c$ in this asymptotic theory
agrees very well with the numerical results.

Our study of the Mott--Hubbard MIT was limited to the Hubbard
model with a non--degenerate orbital. A similar phase transition
should be expected in the Hubbard model with orbital degeneracy.
An important question is, however, whether one can find a physical
system where such a transition is realized. For binary alloys this
might be a very demanding task because the predicted MIT requires
fine tuning of the filling factor with concentration of the alloy
elements as well as special values for the interaction and the
disorder splitting. At present we do not know which alloy system
would be the best candidate for realizing the predicted
Mott--Hubbard MIT.

The most promising candidates for experimental realization may
come from systems with fermions moving on artificial lattices.
Creating a lattice with a binary alloy disorder seem to be
possible either with a matrix of quantum dots with two different
sizes,\cite{dots} or with optical lattices, where
counterpropagating laser beams can be used to trap fermionic atoms.\cite%
{fermions} In the latter case, with proper selection of laser
light and
physical boundaries one can obtain at least quasi--disordered systems\cite%
{optical} where a binary alloy--like distribution is possible.
From the point of view of our theory, where Anderson localization
is not included, quasi--randomness of the optical lattice is not a
major limitation as it yields the effective alloy band splitting
which is crucial for our calculation. Since in the present Mott
insulator the long-range ordering is suppressed due to the
disorder, such a system might be very useful to study the ground
state of, and excited states in, paramagnetic Mott insulators.

\begin{acknowledgments}
We thank R. Bulla for helpful discussions and B. Velick\'y for
valuable correspondence. This work was supported in part by the
SFB 484 of the Deutsche Forschungsgemeinschaft (DFG). Financial
support of KB through KBN-2 P03B 08 224, and of WH through the DFG
is also gratefully acknowledged.\\

\end{acknowledgments}

\appendix*

\section{Calculation of the self--energy}

The self--energy can, in principle, be calculated directly from a Dyson equation: $%
\Sigma(\omega)=G^{-1}(\omega) -\omega +\eta(\omega)$. However,
Bulla {\em et al.}\cite{bulla98} have proposed a different
approach, namely to use a two--particle correlation function to
determine the self--energy. In this way certain systematic errors,
leading to inaccurate spectral weights, are cancelled. In this
Appendix we show how to generalize this method to find the
self--energy in the correlated system with binary alloy disorder.
With a new interpretation of the CPA equation one can use
two--particle correlation functions to determine the
single--particle self--energy. Our method can be applied to an
arbitrary DOS and, as we checked independently, leads to better
convergence of the DMFT equations due to apparent error
cancellations.

Within the DMFT, the disordered Hubbard model is mapped onto a
single--impurity Anderson model which contains the local ionic energy $%
\epsilon _{i}$ as a parameter. For each $\epsilon _{i}$ the
Anderson model is solved independently. It yields the local
$\epsilon _{i}$--dependent Green function $G(\omega ,\epsilon
_{i})=\langle \langle a_{i\sigma }|a_{i\sigma }^{\dagger }\rangle
\rangle $. In addition, for each $\epsilon _{i}$ we can find a
two--particle Green function
\begin{equation}
F(\omega ,\epsilon _{i})=\langle \langle a_{i\sigma }a_{i\bar{\sigma}%
}^{\dagger }a_{i\bar{\sigma}}|a_{i\sigma }^{\dagger }\rangle
\rangle .
\end{equation}%
On the other hand, the equation of motion for $G(\omega ,\epsilon
_{i})$ obtained from the $\epsilon _{i}$--dependent SIAM is
\begin{equation}
(\omega -\epsilon _{i})G(\omega ,\epsilon _{i})-UF(\omega
,\epsilon _{i})-\eta (\omega )G(\omega ,\epsilon _{i})=1,
\end{equation}%
which can be rewritten as
\begin{equation}
\lbrack \omega -\eta (\omega )]G(\omega ,\epsilon
_{i})=1+V_{i}(\omega )G(\omega ,\epsilon _{i}),  \label{a1}
\end{equation}%
where we define a complex and frequency dependent (dynamical)
scattering potential
\begin{equation}
V_{i}(\omega )\equiv \epsilon _{i}+U\frac{F(\omega ,\epsilon
_{i})}{G(\omega ,\epsilon _{i})}.
\end{equation}%
From a formal point of view equation (\ref{a1}) looks like a
single--particle equation for a Green function $G(\omega ,\epsilon
_{i})$ with random potential $V_{i}(\omega )$. The average Green
function $G(\omega )$ is found within the CPA by demanding that
the average of the transfer matrix, given by
\begin{equation}
T_{i}(\omega )=\frac{\epsilon _{i}+U\frac{F(\omega ,\epsilon
_{i})}{G(\omega ,\epsilon _{i})}-\Sigma (\omega )}{1-\left[
\epsilon _{i}+U\frac{F(\omega
,\epsilon _{i})}{G(\omega ,\epsilon _{i})}-\Sigma (\omega )\right] G(\omega )%
}
\end{equation}%
with self--energy $\Sigma (\omega )$, vanishes, i.e., $\int d\epsilon _{i}%
\mathcal{P}(\epsilon _{i})T_{i}(\omega )=0$.

For binary alloy disorder the equation for the self--energy can be
solved analytically, leading to
\begin{equation}
\Sigma (\omega )=\frac{1}{2}\left[ V_{1}(\omega )+V_{2}(\omega )-\frac{1}{%
G(\omega )}\pm \Xi (\omega )\right] ,  \label{a10}
\end{equation}%
where
\begin{widetext}
\begin{eqnarray}
\Xi(\omega)= \left(
\left[V_1(\omega)+V_2(\omega)-\frac{1}{G(\omega)}\right]^2-
\nonumber \right. \left. 4\left[V_1(\omega)  V_2(\omega) -
x\frac{V_1(\omega)}{G(\omega)}
-(1-x)\frac{V_2(\omega)}{G(\omega)}\right] \right) ^{\frac{1}{2}}.
\label{a12}
\end{eqnarray}
\end{widetext}Causality of the Green function requires that the sign in Eq. (%
\ref{a10}) has to be properly chosen: (i) at $\omega \rightarrow -\infty $ ($%
+\infty $) the physical solution of (\ref{a10}) has a $(-)$
$[(+)]$ sign;
(ii) the change of the sign happens an odd number of times at frequencies $%
\omega _{0}$ for which $\mathrm{Im}\Xi (\omega _{0})=0$.



\begin{references}
\bibitem{projection} Explicitely, one finds \\
$\hat{P} |\Psi\rangle = \prod_l \Theta (\Lambda - \epsilon_l +
\Delta /2) \sum A_{l_1\dots l_N}
a^{\dagger}_{l_1\sigma_{l_1}}\cdots a^{\dagger}_{l_N\sigma_{l_N}}
|0\rangle $, where $\Theta$ is a step function and $\Lambda \ll
\Delta$ is a high-energy cutoff.
\end{references}
\end{document}